\documentstyle[procl,epsfig]{article}
\def\Journal#1#2#3#4{{#1} {\bf #2}, #3 (#4)}

\def\PLB{{\em Phys. Lett.}  B}

\def\PRL{\em Phys. Rev. Lett.}

\def\PRD{{\em Phys. Rev.} D}

\def\ZPC{{\em Z. Phys.} C}

\def\be{\begin{equation}}

\def\ee{\end{equation}}

\def\bea{\begin{eqnarray}}

\def\eea{\end{eqnarray}}

\begin{document}

\begin{flushright}
\vspace*{-3.2cm}
ANL-HEP-CP-96-50 \\
RAL-TR-96-045 \\
June 1996 \\ 
\vspace*{0.9cm}
\end{flushright}

\title{ISOLATED PROMPT PHOTON PRODUCTION AT HERA*}

\author{ L.E. GORDON }

\address{High Energy Physics Division, Argonne National Laboratory, \\
Argonne IL 60439, USA}

\author{ W. VOGELSANG }

\address{Rutherford Appleton Laboratory,  
Chilton DIDCOT, Oxon OX11 0QX, England}

\maketitle\abstracts{
We study the expectations for prompt photon production rates at HERA in 
a fully consistent next-to-leading order (NLO) QCD analysis, 
taking into account the effects of experimental isolation 
requirements. In particular we examine the sensitivity of the isolated 
cross section to the photon's gluon content.}

The utility of hadronic large-$p_T$ prompt photon production for providing 
constraints on the proton's gluon content has led to the suggestion that 
it may also prove useful in determining the photonic gluon distribution at 
HERA via the study of photoproduction of direct photons~\cite{aur1}. 
The process was subsequently studied quite extensively, with NLO QCD
corrections to it partly taken into account~\cite{aur1,bawa,aur2,gs1}. 
The main shortcoming of all these studies was    
that only the fully inclusive cross section was considered. Since, 
however, HERA is a collider it will be necessary to perform isolation 
cuts in the experiments in order to unambiguously identify the photon 
signal from the huge hadronic background, just as it is necessary at existing 
hadron colliders~\cite{cdf}. Indeed, first reports on HERA prompt photon 
results~\cite{dull,bus} confirm this view. In order
to make reliable and meaningful predictions it is crucial that the
theoretical calculation correctly includes the experimental
isolation constraints. This was done in~\cite{gv} where the first 
complete NLO analysis, at the same time fully taking into account the effects 
of isolation, was presented. In the following we provide an update of our 
study~\cite{gv}, for which we believe it is time now since most sets of parton 
distributions used in~\cite{gv} have been updated in the 
meantime~\cite{grv,mrs,cteq,gs}.
Furthermore, we will also more closely match cuts used in other
HERA photoproduction experiments performed so far~\cite{zeus}, 
hereby making our predictions more realistic and more directly comparable
to future data. 

As with all photoproduction processes at HERA, there are two types of 
contributions to the cross section, the so-called direct and resolved 
contributions. In the case of prompt photon production 
there are two further subclasses in each category, which we label the 
fragmentation and non-fragmentation processes according to whether the prompt
photon is produced directly in the hard scattering process or via 
fragmentation of a final state 
parton\footnote{* Invited talk presented by W. Vogelsang at the
'Int. Workshop on Deep Inelastic Scattering and Related Phenomena',
Rome, Italy, April 15-19, 1996.}.

The cross section for $ep \rightarrow \gamma X$ can thus be schematically 
written as
\begin{equation}  \label{cross}
d\sigma_{ep}^{\gamma} = \sum_{f_p,f_e,f} \int_{\otimes} f_{p} (x_p,M^2) f_{e} 
(x_e,M^2) d\hat{\sigma}_{f_p f_e}^f (p_{\gamma},x_p,x_e,z,M^2,M_F^2) 
D_f^{\gamma}(z,M_F^2) \: ,
\end{equation} 
where $f_p = q,\bar{q},g$ and $f_e,f=q,\bar{q},g,\gamma$. The 'electron 
structure function' $f_e (x_e,M^2)$ at scale $M$ is defined by the convolution
\begin{equation}  \label{esf}
f_{e} (x_e,M^2) =  \int^1_{x_e} \frac{dy}{y} P_{\gamma /e}(y)
f^{\gamma}\left( \frac{x_e}{y},M^2\right) \:\:\: ,
\end{equation} 
where $f^{\gamma}$ denotes the corresponding {\em photonic} parton 
distribution. The definition is readily extended to the direct case 
('$f_e=\gamma$') by replacing $f_e (x_e/y,M^2) \rightarrow \delta (1-x_e/y)$.
The flux of quasi-real photons radiated from the electron beam is estimated
in the Weizs\"{a}cker-Williams approximation,
\begin{equation}  \label{weiz}
P_{\gamma/e}(y)=\frac{\alpha_{em}}{2\pi}\left[ \frac{1+(1-y)^2}{y} \right] 
\ln \frac{Q_{max}^2 (1-y)}{m_e^2 y^2} \:\:\: ,
\end{equation}
where $m_e$ is the electron mass.
As in~\cite{zeus} we will use an upper cut $Q_{max}^2=4$
GeV$^2$ on the virtuality of the incoming photon, and we will also 
apply the $y$-cut $0.2 \leq y \leq 0.8$ which will serve to bring our
predictions closer to the actual experimental situation than in our 
previous study~\cite{gv}. 
Finally, in Eq.~(\ref{cross}) $D^{\gamma}_f (z,M_F^2)$ is the fragmentation 
function at scale $M_F$ for the fragmentation of parton $f$ into a photon. 
We include in its definition the non-fragmentation case ('$f=\gamma$')
where $D^{\gamma}_f (z,M_F^2) \rightarrow \delta (1-z)$.
  
The isolation technique used at hadron colliders~\cite{cdf} is to define 
a cone centered on the photon with radius $R=\sqrt{\Delta \phi^2+\Delta 
\eta^2}$, inside which the allowed amount of hadronic energy 
is restricted to be below $\epsilon E_{\gamma}$ with the prompt photon's
energy $E_{\gamma}$ and $\epsilon \sim {\cal O}
(0.1)$. A cone isolation method has also been applied in 
the first experimental prompt photon studies at HERA~\cite{dull,bus}. 
In order to take into account the effects of isolation on the cross section 
(\ref{cross}) in NLO we use the analytical method 
of~\cite{wir1} which was extended to the HERA situation in~\cite{gv}.

To see whether prompt photon production at HERA will give useful information on 
the parton, in particular the gluon, distributions of the photon
we will compare the predictions obtained for two different 
NLO sets of photonic parton densities suggested in the 
literature, the GRV~\cite{grvgam} and GS\cite{gs} (recent update) sets.
On the proton side, we provide a comparison of the results obtained using
the latest GRV(94)~\cite{grv}, MRS(${\rm A'}$)~\cite{mrs} and 
CTEQ(3M)~\cite{cteq} versions. 
All other ingredients and parameters for our analysis 
are chosen exactly as in our previous study~\cite{gv}, with the
exception of the Weizs\"{a}cker-Williams spectrum (see (\ref{weiz})). 
In particular, we choose the isolation parameters $\epsilon=0.1$, $R=0.4$.

In Fig.~1(a) we show the fully inclusive and the isolated cross sections 
vs. the prompt photon's rapidity $\eta$ at $p_T=5$ GeV, using GRV distributions 
throughout. We also illustrate the expected strong reduction of the 
fragmentation contribution (i.e., the sum of direct fragmentation and resolved 
fragmentation) due to isolation. 
\begin{figure}[h]
\vspace*{-0.6cm}
\hspace*{0.1cm}
\epsfig{file=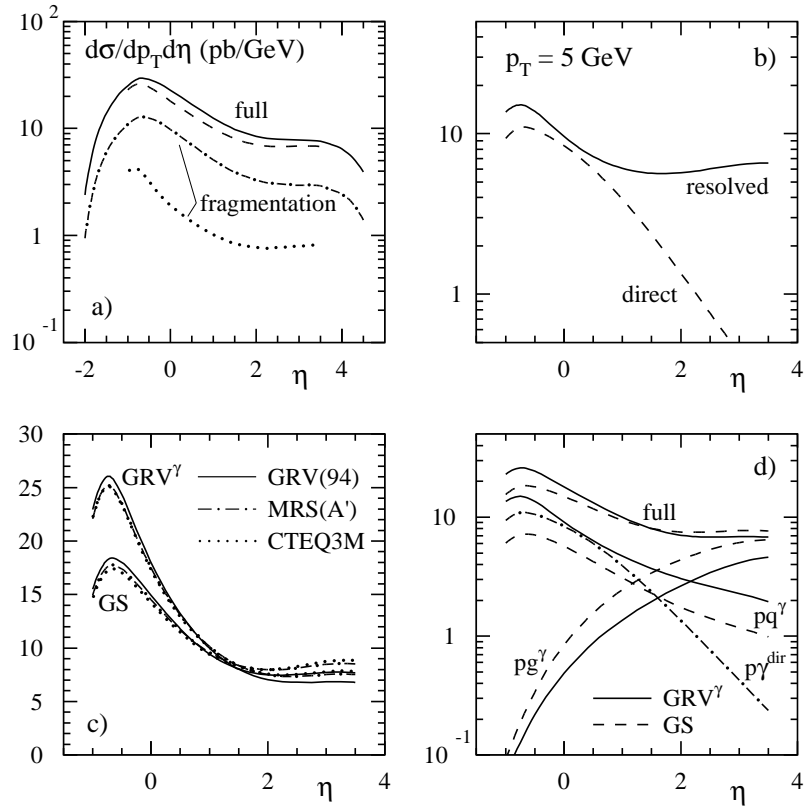,width=12.2cm,angle=0}
\vspace*{-1.1cm}
\caption{{\bf (a)} Comparison of fully inclusive and isolated 
results for the full cross section and its fragmentation 
part. {\bf (b)} Resolved and direct contributions to the isolated 
cross section. {\bf (c)} Full isolated cross section for various 
sets of parton distributions of the proton and the photon.
{\bf (d)} Full isolated cross section and its decomposition according
to Eq.~(4) for GRV and GS photonic parton distributions.} 
\end{figure}  
Fig.~1(b) compares the corresponding resolved 
and direct contributions to the isolated cross section, both including
their non-fragmentation as well as their fragmentation parts.
The direct contribution is strongly peaked at negative rapidities, 
corresponding to the probing of the proton at small $x_p$ by an energetic 
photon. The resolved contribution remains sizeable and dominant 
also at positive $\eta$. It has two peaks: The one at $\eta \approx 3$ 
corresponds to the probing of the photon distributions at rather 
small $x_{\gamma}$, and the cross section is expected to be sensitive to 
$g^{\gamma}$ here. The somewhat larger peak around 
$\eta=-1$ is due to the probing of the protonic gluon distribution at 
small $x_p$ by the photonic quark distributions at large $x_{\gamma}$. 

In Fig.~1(c) we study the sensitivity of the cross section to the proton and 
photon structure functions. It becomes obvious that there is a significant 
difference between the predictions given by the GRV and GS photonic parton 
distributions at negative $\eta$, where the uncertainties coming from the 
proton structure functions are rather small. To further analyze this 
issue, Fig.~1(d) shows the decomposition into the contributions of the 
subprocesses~\cite{gv} 
\begin{equation}  \label{decomp}
p g^{\gamma}\rightarrow \gamma X \: , \:\:   
p q^{\gamma}\rightarrow \gamma X \: , \:\:   
p \gamma^{dir} \rightarrow \gamma X  \: , \:\:    
\end{equation}
using a fixed set (GRV) of proton structure functions. It becomes 
clear that differences between the photonic quark and gluon distributions
in the two sets~\cite{grvgam,gs} show up rather strongly, but that they 
partly compensate in the sum. As expected, the processes involving 
$g^{\gamma}$ dominate the cross section at large positive $\eta$.
On the other hand, as Fig.~1(c) shows, the present uncertainties stemming 
from the proton's parton (in particular $u$ quark) distributions will somewhat 
obscure the differences between the results for the two $g^{\gamma}$ here. 

\end{document}